\documentstyle[11pt,paspconf,epsf,rotate]{article}

\begin{document}
 
\title{ Discs and Planetary Formation}

\author{J. C. B. Papaloizou}
\affil{Astronomy Unit,  Queen Mary \& Westfield College \\ 
Mile End Road, London~E14NS, UK}
 
\author{C. Terquem}
\affil{Lick Observatory, University of California \\ 
Santa Cruz, CA~95064, USA}

\author{R. P. Nelson}
\affil{Astronomy Unit, Queen Mary \& Westfield College \\ 
Mile End Road, London~E14NS, UK}

\begin{abstract}
The formation, structure and evolution of protoplanetary discs is
considered.  The formation of giant planets within the environment of
these models is also discussed.
\end{abstract}
 
 
 

\section{Introduction}

\noindent 
The hypothesis that the planets in the solar system were formed in a
flattened differentially rotating gaseous disc was originally proposed
by Laplace (1796).  The suggestion arises naturally in view of their
orbital configuration. They are in near circular orbits which
approximately lie in the same plane.  Models in which planet formation
occurs in gaseous discs have been the subject of much theoretical
interest in recent times (e.g. Lin \& Papaloizou 1985, 1993 and
references therein).

\noindent When discussing the large scale physical properties of such
a disc, it is convenient to make the idealization that it is
infinitesimally thin or completely flat. 
When, as
in the present solar system, most of the mass lies in a central star
of mass $M_{*},$ the local gravitational acceleration can be taken to be
 $g= GM_{*}/r^2.$ A gaseous element at radius
$r$ then rotates in circular orbit with angular velocity $\Omega,$ 
 which, to a good approximation, is obtained from
centrifugal balance by equating $g=r\Omega^2.$ This gives Kepler's law
\begin{equation} 
\Omega^2 ={GM_{*}\over r^3}.
\end{equation}
 In this approximation,  in which  forces other than that due to the gravity of
the central star are neglected,   material rotates 
 conserving its specific  angular momentum $j=r^2\Omega.$ 
This then   changes only slowly  when viscous
or other perturbative forces are taken account of.

\subsection{Observations of protostellar discs}

\noindent Between 25 and 75 percent of young stellar objects in the
Orion nebula appear to have discs (McCaughrean \& Stauffer 1994).
Masses of $\sim 10^{-2\pm 1}$~M$_{\odot},$ and dimensions $\sim 40 \pm
20$~AU have been estimated (Beckwith \& Sargent 1996).

\noindent The presence of discs on the scale of astronomical units has
been inferred from the infrared excesses observed in about half of all
T~Tauri stars. Their colors and magnitudes can be fitted by those
expected for young pre--main sequence stars with ages $ \sim 10^6$~yr
(Strom et al. 1993). The infrared emission may be produced by the
gravitational potential energy liberated by matter flowing inwards at
a rate $\dot M \sim 10^{-8 \pm 1}$~M$_\odot$~yr$^{-1}.$ The non
observation of discs around older T~Tauri stars together with these
values of $\dot M$ suggest a disc lifetime of $\sim 10^7$~yr.

\subsection{Formation of discs}

It is believed that protostars and protoplanetary discs derive from
interstellar matter contained in molecular clouds.  Observations
(Goodman et al. 1993) indicate that typical star--forming dense cores
in dark molecular clouds have specific angular momentum $j >
6\times10^{20}$~cm$^2$~s$^{-1}$. When these clouds undergo
gravitational collapse, $j$ is initially approximately conserved
because the collapse is dynamical.  Gas in the outer parts will not be
able to fall directly to the centre ($r=0$) if $j$ is not zero.

\noindent To obtain an estimate of the initial size of the disc, we
consider the idealized situation when the pre--collapse cloud is a
cold rotating sphere of mass $M$ and radius $R,$ with a single axis of
rotation. Matter located on the rotation axis has no angular momentum
so it can fall directly to the centre. An estimate of the time
required, $t_{ff},$ is given by the time required to free fall from
rest through a distance $R$ under the initial inward gravitational
acceleration at the surface, $g=GM/R^2.$ This gives $t_{ff} =
\sqrt{2R/g}= \sqrt{2R^3/(GM)}.$ In terms of the
initial mean density ${\overline \rho}$ of the cloud,  $t_{ff} =
\sqrt{ 3/(2\pi G {\overline \rho} )}.$ For a cloud core with
$M= 1$~M$_{\odot}$ and $R=0.1$~pc, this gives  $t_{ff} \sim
6\times10^{5}$~yr.

\noindent  While matter located on the rotation axis
can move directly to the centre, matter in the equatorial plane at
the surface of the sphere will be at the outermost radius $R_d$ of a
disc after collapse. Conservation of
specific angular momentum determines $R_d$. Assuming the total mass, $M,$ to be
concentrated at the centre, the angular velocity at the outer edge of
the disc will be given by Kepler's law such that
\begin{equation}
\Omega^2 = {j^2\over R_d^4}={GM\over R_d^3}.
\end{equation}
Thus $R_d$ is given in terms of the conserved quantities $j$ and $M$
by $R_d = j^2/(GM).$ Adopting $j = 6\times 10^{20}$~cm$^2$~s$^{-1}$ and
$M=1$~M$_{\odot},$ we find that $R_d\sim 180$~AU. The characteristic
dimension of such a disc is about an order of magnitude larger than
our present solar system but is similar to those of protostellar discs
now being observed by direct imaging.

\section{Early evolution }

The formation of a protostellar disc through the collapse of a
molecular cloud core takes $10^5$--$10^6$~yr.  During the early stages
when the disc is still embedded (class~0/1 object) and has a
significant mass compared to the central star, there may exist strong
disc winds and bipolar outflows (e.g. Reipurth et al. 1997) with
associated magnetic fields.  During this stage a hydromagnetic disc
wind may be an important means of angular momentum removal for the
system (see Papaloizou \& Lin 1995, and references therein).

When the mass of the disc is significant compared to that of the star,
there may be a short period ($\sim 10^5$~yr) of non axisymmetric
global gravitational instability with associated outward angular
momentum transport (Papaloizou \& Savonije 1991; Heemskerk et
al. 1992; Laughlin \& Bodenheimer 1994; Pickett et al. 1998) that
results in additional mass growth of the central star.  This
redistribution may occur on the dynamical timescale (a few orbits) of
the outer part of the disc and so may be quite rapid, on the order of
$10^5$~yr for $R=500$~AU.  The parameter governing the importance of
disc self--gravity is the Toomre parameter, $Q= M_{*} H/(M_d r),$ with
$M_d$ being the disc mass contained within radius $r$ and $H$ being
the disc semi--thickness.  In this review we shall take $H$ to be the
distance between the disc mid--plane and surface.  This is usually a
factor of 2--3 larger than the vertical extent reached by a disc
particle moving through the disc mid--plane with the local sound speed
$c_s.$ Thus $ H \sim (2 c_s)/\Omega.$ Typically $H/r \sim 0.1$
(Stapelfeldt et al. 1998) such that the condition for the importance
of self--gravity, $Q \sim 1,$ gives $M_d \sim 0.1 M_{*}.$

The characteristic scale associated with growing density perturbations
in a disc undergoing gravitational instability with $Q\sim 1$ is $\sim
H,$ and the corresponding mass scale is $M_d(H/r)^2 \sim
M_{*}(H/r)^3,$ which is $\sim 1$~M$_J$ for $H/r\sim 0.1$ and $M_{*}=
1$~M$_{\odot},$ with M$_J$ being Jupiter's mass.  Gravitational
instability does not necessarily lead to fragmentation (e.g. Pickett
et al. 1998), nonetheless it has been proposed as a mechanism for
directly forming giant planets by Cameron (1978) and more recently by
Boss (1998).

During the period of global gravitational instability, it is
reasonable to suppose that the disc mass is quickly redistributed and
reduced such that global gravitational stability is restored ($Q >
1$), after which further disc evolution occurs on a longer timescale
governed by viscosity with effects due to self--gravity being small.

\subsection {Viscous evolution}

During this phase, the disc may attain a configuration similar to that
expected for the minimum mass primordial solar nebula, $M_d \sim
10^{-2}$--$10^{-1}$~M$_{\odot}. $ Planets have been proposed to form
out of such a disc by a process of growth through planetesimal
accumulation followed, in the giant planet case, by gas accretion
(Safronov 1969; Wetherill \& Stewart 1989).

During this evolutionary phase,
 it is reasonable to regard the disc as an axisymmetric
configuration in which, to a first approximation, material orbits in
circlar Keplerian orbits. 
 However, other weaker forces due to internal pressure,
viscosity, or magnetic fields may also operate in the disc. These
can result in   angular momentum redistribution  on a long timescale.  In
order to flow inwards, material has to transport its angular momentum
outwards to matter at larger radii.  The angular momentum transport
process determines the timescale on which mass accretion can occur and
hence the evolutionary timescale of the disc.

\noindent Historically, the first angular momentum transport mechanism
to be considered was through the action of viscosity (von
Weizs\"acker 1948).  This acts through the friction of neighbouring
sections of the disc upon each other. The inner regions rotate faster
than the outer regions and thus viscous friction tends to communicate
angular momentum from the inner parts of the disc to the outer
parts. In order to result in evolution on astronomically interesting
timescales, it is necessary to suppose that an anomalously large
viscosity is produced through the action of some sort of
turbulence. The magnitude of the viscosity is usually parameterized
through the Shakura \& Sunyaev (1973) $\alpha$ prescription.

\noindent
In this we suppose the kinematic viscosity coefficient $\nu = \alpha
c_s^2/\Omega \sim \alpha H^2\Omega, $ where $\alpha$ is a
dimensionless constant which must be less than unity.  Currently the
most likely mechanism for producing turbulence is through
hydromagnetic instabilities (Balbus \& Hawley 1991) which might
produce $\alpha \sim 0.01,$ provided the disc has adequate ionization.

Gammie (1996) has proposed a model in which viscosity only operates in
the surface layers where external sources of ionization such as cosmic
rays can penetrate.  Such layered models may have an interior dead
zone of material for $r > 0.1$~AU and, although they can be considered
as models with a variable $\alpha,$ they will behave somewhat
differently from the standard models considered here which have
$\alpha$ assumed to be constant throughout.

\section{ The diffusion equation for disc evolution}

\noindent The evolution of a viscous disc is controlled by angular
momentum conservation. The conservation of specific angular momentum
may be expressed in the form ( eg. Papaloizou \& Lin 1995)

\begin{equation}
\rho {Dj\over Dt}=-\nabla\cdot {\bf F} =- {1\over r} {\partial \over
\partial r}(r F_r)-{\partial \over \partial z}( F_z),
\label{cons}
\end{equation}
where ${\bf F}= (F_r,F_z)$ is the angular momentum flux, with the
vertical coordinate  being denoted by $z$, and $\rho$ is the mass density.

\noindent The process of vertical averaging applied to
equation~(\ref{cons}) in the case of a Keplerian disc yields

\begin{equation}
\Sigma \langle v_r\rangle {d j\over d r} =-{1\over r}{\partial \over
\partial r}(r\langle F_r\rangle),
\label{ave}
\end{equation} 
where $v_r$ is the radial velocity in the disc and $\Sigma
=\int^{H}_{-H}\rho dz$ is the surface mass density. The vertical
average for a quantity ${\cal Q}$ is defined by
\begin{equation} 
\langle {\cal Q}\rangle = {\int^{H}_{-H}\rho {\cal Q}
 dz\over \int^{H}_{-H}\rho dz}.
\end{equation}

\noindent We assume ${\bf F}$ vanishes on the disc boundaries 
ensuring that the total angular momentum of the disc is conserved.
We comment that this implies that angular momentum loss through
winds or gain through mass infall is neglected.
The radial angular momentum flux arising from
viscosity is given by

\begin{equation} 
F_r =-r^2\rho \nu{d\Omega\over dr},
\label{fl}
\end{equation} 

\noindent and thus
 
\begin{equation} 
\langle F_r \rangle = -r^2\langle \nu\rangle\Sigma {d\Omega\over dr},
\label{flu}
\end{equation}

\noindent   For a
Keplerian disc the angular velocity decreases outwards so that the
angular momentum flux is directed outwards as required in order that
mass accretion may occur.

\noindent Using equation~(\ref{flu}) in equation~(\ref{ave}) and
solving for $\langle v_r\rangle $ gives
$$\langle v_r\rangle ={1\over r\Sigma }\left({d j\over d
r}\right)^{-1} {\partial \over \partial r} \left(r^3\langle
\nu\rangle\Sigma {d\Omega\over dr}\right).$$ For a Keplerian disc in
which  $j=\sqrt{GM_{*}r},$ we get

\begin{equation}
\langle v_r\rangle =-{3\over r^{1/2}\Sigma } {\partial \over \partial
r}\left(r^{1/2}\langle \nu\rangle\Sigma \right). \label{vrad}
\end{equation}

\noindent Thus the radial velocity in the disc is determined in terms
of the kinematic viscosity. For a steady state disc in which $\langle
\nu\rangle \Sigma$ is constant, we find $$\langle v_r\rangle
=-{3\langle \nu\rangle \over 2 r}.$$ This velocity is negative,
consistent with accretion onto the central object.

\noindent To obtain   a general equation   governing the disc, the
velocity~(\ref{vrad}) is  used together with the vertically averaged
continuity equation
$${\partial \Sigma \over \partial t} + {1 \over r} {\partial \over
\partial r} \left(\Sigma r\langle v_r\rangle\right) =0.$$

\noindent The global evolution of the disc is thus found to be
governed by a single diffusion equation for the surface density which
takes the form (Lynden--Bell \& Pringle 1974)

\begin{equation} 
{\partial \Sigma \over \partial t} = {3 \over r}{\partial \over
\partial r}\left(r^{1/2}{\partial \over \partial r} \left( \Sigma
\langle \nu \rangle r^{1/2}\right) \right)
\label{aa4} .
\end{equation}

\noindent According to this the characteristic evolution timescale for
the disc will be the global diffusion timescale. The diffusion
coefficient is $3\langle \nu \rangle$ and, adopting a characteristic
radius $r,$ the  global diffusion timescale is
$$t_{e} = {r^2\over 3\langle \nu \rangle}.$$ 

\noindent  
For viscosity coefficient $\nu,$  parameterized through the
$\alpha$ prescription  so that $\nu = \alpha c_s ^2 / \Omega \sim
\alpha H^2\Omega, $ we obtain 
$$t_e \Omega =(1/3)(r/H)^2\alpha^{-1}.$$

\noindent For a protostellar disc of size $100$~AU with $H/r\sim 0.1,$
and a central solar mass, we find $t_e =5 \times 10^5 (0.01/\alpha)
$~yr. This gives lifetimes comparable to those estimated for discs
around T Tauri stars if $\alpha \sim 10^{-3}$--$10^{-2}.$

Equation~(\ref{aa4}) enables the evolution of the disc to be
determined if $\langle \nu \rangle $ is specified as a function of
$\Sigma.$ This can be done by  solving for the vertical structure of
the disc.

\section{Vertical structure calculations}
\label{sec:disk_models}

\subsection{Basic equations}

We  adopt the equation of vertical hydrostatic equilibrium:

\begin{equation}
\frac{1}{\rho} \frac{\partial P}{\partial z} = - \Omega^2 z ,
\label{dPdz}
\end{equation}

\noindent and the energy equation, which
states that the rate of energy  removal by radiation is locally
balanced by the rate of energy production by viscous dissipation:

\begin{equation}
\frac{\partial {\cal F}}{\partial z} = \frac{9}{4} \rho \nu \Omega^2 ,
\label{dFdz1}
\end{equation}

\noindent where ${\cal F}$ is the radiative flux of energy through a
surface of constant $z$  which is given by:

\begin{equation}
{\cal F} = \frac{- 16 \sigma T^3}{3 \kappa \rho}
\frac{\partial T}{\partial z} .
\label{dTdz}
\end{equation}

\noindent Here $P$ is the pressure, $T$ is the temperature, $\kappa$
is the opacity, which in general depends on both $\rho$ and $T$, and
$\sigma$ is the Stefan--Boltzmann constant.

\noindent To close the system of equations, we relate $P$, $\rho$ and
$T$ through the equation of state of an ideal gas: 

\begin{equation}
P = \frac{\rho k T}{\mu m_H} ,
\label{state}
\end{equation}

\noindent where $k$ is the Boltzmann constant, $\mu$ is the mean
molecular weight and $m_H$ is the mass of the hydrogen atom. Since the
main component of protostellar disks at the temperatures we consider
is molecular hydrogen, we take $\mu=2$. We denote the isothermal sound
speed by $c_s$ ($c_s^2=P/\rho$).

As above, we adopt the $\alpha$--parametrization of Shakura \& Sunyaev
(1973), so that the kinematic viscosity is written $\nu=\alpha
c_s^2/\Omega$. In general, $\alpha$ is a function of both $r$ and
$z$. However, we shall limit our calculations presented below to cases
with constant $\alpha$. With this formalism, equation~(\ref{dFdz1})
becomes:

\begin{equation}
\frac{\partial {\cal F}}{\partial z} = \frac{9}{4} \alpha \Omega P .
\label{dFdz}
\end{equation}

\subsubsection{Boundary conditions} 

We have to solve three first order ordinary differential
equations for the three variables ${\cal F}$, $P$ (or equivalently
$\rho$), and $T$ as a function of $z$ at a given radius $r.$  Accordingly,
we need
three boundary conditions at each $r$. We shall denote with a
subscript $s$ values at the disk surface.

  The first boundary condition is obtained by integrating
equation~(\ref{dFdz1}) over $z$ between $-H$ and $H$. Since by
symmetry ${\cal F}(z=0)=0$, this gives:

\begin{equation}
{\cal F}_s = \frac{3}{8 \pi} \dot{M}_{st} \Omega^2 ,
\label{Fs}
\end{equation}

\noindent where we have defined $\dot{M}_{st} = 3 \pi
\langle{\nu}\rangle \Sigma.$ If the disk were in a steady state,
$\dot{M}_{st}$ would not vary with $r$ and would be the constant
accretion rate through the disk. In general however, this quantity
does depend on $r$.

The  second boundary condition is determined by  the fact that very
close to the surface of the disc, since the optical depth $\tau_{ab}$
above the disc is small, we have:

$$ P_s = g_s\tau_{ab}/{\kappa_s}.$$

This condition is familiar in stellar structure (e.g. Schwarzschild
1958).  Using $g_s =\Omega^2 H,$ we thus obtain
\begin{equation}
P_s = \frac{\Omega^2 H \tau_{ab}}{\kappa_s} .
\label{Ps}
\end{equation}
Provided $\tau_{ab} \ll 1,$ the results do not
depend on the value of $\tau_{ab}$ we choose (see below).

The third boundary condition we use is 

\begin{equation}
2 \sigma \left( T_s^4 - T_b^4 \right) - \frac{9 \alpha k T_s \Omega}{8
\mu m_H \kappa_s} - \frac{3}{8 \pi} \dot{M}_{st} \Omega^2 = 0 .
\label{Ts}
\end{equation}
Here the disc is assumed immersed in a medium with background
 temperature $T_b.$ The surface opacity $\kappa_s$ in general depends
 on $T_s$ and $\rho_s$ and we have used $c_s^2=kT/(\mu m_H)$.

The boundary condition~(\ref{Ts}) is the same as that used by
Levermore \& Pomraning (1981) in the Eddington approximation (their
eq.~[56] with $\gamma=1/2$).  In the simple case when $T_b=0,$ and the
surface dissipation term involving $\alpha$ is set to zero, it simply
relates the disc surface temperature to the emergent radiation flux.

\subsubsection{ Disc models}

At a given radius $r$ and for a given values of the parameters
$\dot{M}_{st}$ and $\alpha$, we solve equations~(\ref{dPdz}),
(\ref{dTdz}) and~(\ref{dFdz}) with the boundary conditions~(\ref{Fs}),
(\ref{Ps}) and~(\ref{Ts}) to find the dependence of the state
variables on $z.$ The opacity is taken from Bell \& Lin (1994). This
has contributions from dust grains, molecules, atoms and ions.  It is
written in the form $\kappa=\kappa_i \rho^a T^b$ where $\kappa_i$, $a$
and $b$ vary with temperature.

\noindent The equations are integrated using a fifth--order
Runge--Kutta method with adaptive step length (Press et al. 1992).
For a specified $\dot{M}_{st},$ we determine the value of $H,$ the
vertical height of the disc surface.  This is done iteratively.
Starting from an estimated value of $H,$ after satisfying the surface
boundary conditions, the equations are integrated down to the
mid--plane $z=0.$ The condition that ${\cal F}=0$ at $z=0$ will not in
general be satisfied.  An iteration procedure (e.g. the
Newton--Raphson method) is used to adjust value of $H$ until ${\cal
F}=0$ at $z=0$ to a specified accuracy.

An important point to note is that as well as finding the disc structure,
 we also determine the surface density $\Sigma$ for a given $\dot{M}_{st}
=3\pi\langle \nu \rangle \Sigma.$ Thus a $\langle \nu \rangle, \Sigma$
relation is derived.

In the calculations presented here, we have taken the optical depth of
the atmosphere above the disk surface $\tau_{ab}=10^{-2}$ and a
background temperature $T_b=10$~K.  In these calculations  the
temperatures are  lower than about 4,000~K, so that, at the densities we
consider, hydrogen is not dissociated and the mean molecular weight
$\mu=2$.

\noindent In the optically thick regions of the disk, the  value
of $H$  is independent of the value of $\tau_{ab}$ we choose. However,
this is not the case in optically thin regions where we find that, as
expected, the smaller $\tau_{ab},$ the larger $H.$ However, this
dependence of $H$ on $\tau_{ab}$ has no physical significance, since
in all cases, the mass is concentrated towards the disk mid--plane in
a layer with thickness independent of $\tau_{ab}.$

\noindent For example, for
$\dot{M}_{st}=10^{-8}$~M$_{\odot}$~yr$^{-1}$ and $\alpha=10^{-2}$, we
find, at $r=100$~AU, that $H/r=0.08$ and 0.24 for $\tau_{ab}=10^{-2}$
and $10^{-5}$, respectively. However, in both cases, the surface
density, the optical thickness and the mid--plane temperature are the
same. Only the mid--plane pressure varies slightly (by about 30\%)
between these cases.

\begin{figure}
\plotone{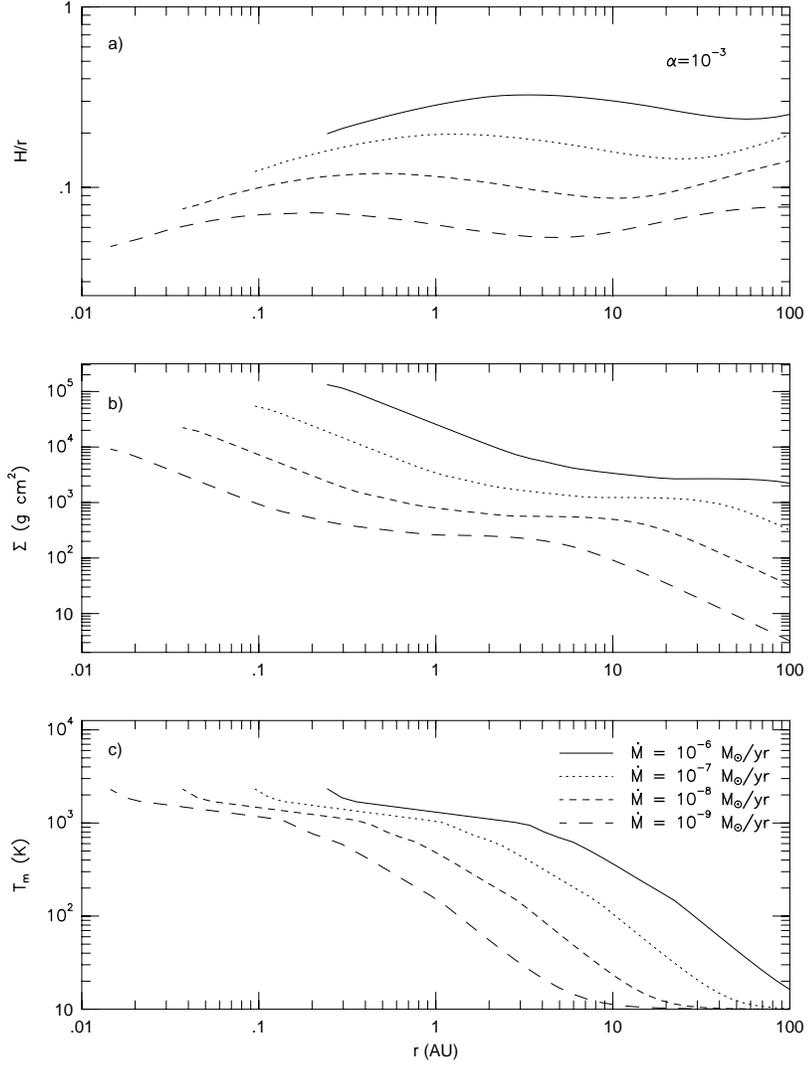}
\caption[]{Shown is $H/r$ ({\it plot~a}), $\Sigma$ in units
g~cm$^{-2}$ ({\it plot~b}) and $T_m$ in K ({\it plot~c}) vs. $r/{\rm
AU}$ using a logarithmic scale for $\dot{M}_{st}$~(in units
M$_{\odot}$~yr$^{-1}$)~$=10^{-6}$ ({\it solid line}), $10^{-7}$ ({\it
dotted line}), $10^{-8}$ ({\it short--dashed line}) and $10^{-9}$
({\it long--dashed line}) and for $\alpha=10^{-3}$.}
\label{fig1}
\end{figure}

In Figures~\ref{fig1}a--c, we plot $H/r$, $\Sigma$ and the mid--plane
temperature $T_m$ versus $r$ for $\dot{M}_{st}$ between $10^{-9}$ and
$10^{-6}$~M$_{\odot}$/year (assuming this quantity is the same at all
radii, i.e. the disk is in a steady state) and for illustrative
purposes we have adopted  $\alpha=10^{-3}$.

Figure~\ref{fig1}a indicates that the outer parts of the disk are
shielded from the radiation of the central star by the inner
parts. This is in agreement with the results of Lin \& Papaloizou
(1980) and Bell et al. (1997). For $\alpha=10^{-3}$, the radius beyond
which the disk is not illuminated by the central star varies from
0.2~AU to about 3~AU when $\dot{M}_{st}$ goes from $10^{-9}$ to
$10^{-6}$~M$_{\odot}$~yr$^{-1}$. These values of the radius move to
0.1 and 2~AU when $\alpha=10^{-2}$.  Since reprocessing of the stellar
radiation by the disk is not an important heating factor below these
radii, this process will in general not be important in protostellar
disks. We note that this result is independent of the value of
$\tau_{ab}$ we have taken. Indeed, as we pointed out above, only the
thickness of the optically thin parts of the disk (which do not
reprocess any radiation) gets larger when $\tau_{ab}$ is decreased.

The values of $H/r$, $\Sigma$ and $T_m$ we get are qualitatively
similar to those obtained by Lin \& Papaloizou (1980), who adopted a
prescription for viscosity based on convection, and Bell et al.
(1997). Our values of $H/r$ are somewhat larger though, since $H$ is
measured from the disk mid--plane to the surface such that $\tau_{ab}$
is small, and not 2/3 as usually assumed.  However, as we commented
above, this has no effect on the other physical quantities. We also
recall that $H,$ as defined here, is about 2--3 times larger than
$c_s/\Omega,$ with $c_s$ being the mid--plane sound speed, which is
often used to define the disc semi--thickness.

\subsubsection{Time dependent evolution of the disc}

We determine the evolution of the radial structure of a non--steady
$\alpha$--disk by solving the diffusion equation (\ref{aa4}). To do
this, we need to use the relation between $\dot{M}_{st}=3 \pi \langle
\nu \rangle \Sigma$ and $\Sigma$ at each radius. Interpolation of or
piece--wise power law fits to numerical data may be used to represent
this relation and more details of these will be published elsewhere.
We note that they can be used either to compute $\Sigma$ from
$\dot{M}_{st}$ or $\dot{M}_{st}$ from $\Sigma.$

\begin{figure}
\plotone{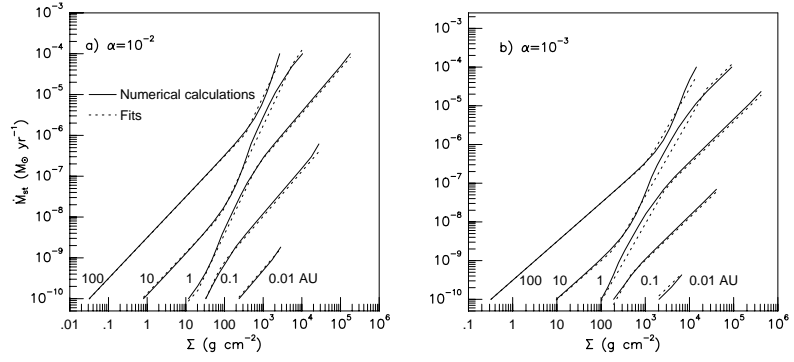}
\caption[]{$\dot{M}_{st}$ in units M$_{\odot}$~yr$^{-1}$ vs. $\Sigma$
in units g~cm$^{-2}$ using a logarithmic scale for $\alpha=10^{-2}$
({\it plot a}) and $10^{-3}$ ({\it plot b}). Both the curves
corresponding to the numerical calculations ({\it solid line}) and the
fits ({\it dashed line}) are shown. The label on the curves represents
the radius, which varies between 0.01 and 100~AU.}
\label{fig2}
\end{figure}

In Figures~\ref{fig2}a--b we plot both the curves $\dot{M}_{st} \left
( \Sigma \right)$ that we get from the vertical structure integrations
as described above and those obtained from piece--wise power law fits.
Figures~\ref{fig2}a and~\ref{fig2}b are for $\alpha=10^{-2}$ and
$10^{-3}$, respectively. In each case the radius varies between 0.01
and 100~AU.  If we calculate $\dot{M}_{st}$ using the fits with
$\Sigma$ as an input parameter, the average error is 22, 18, 13\% and
the maximum error is 55, 48, 42\% for $\alpha= 10^{-4}, 10^{-3},$ and
$10^{-2}$, respectively.  We see that the fits give a good
approximation.

Using the $\left( \dot{M}_{st}, \Sigma \right)$ relation derived from
the integrations, we solved equation~(\ref{aa4}) using explicit finite
difference techniques.  We considered the situation of a disc with
initially $0.1$~M$_{\odot}$ for which $\Sigma \propto r^{-1},$
extending to 100~AU. The central star had $M_{*} =1$~M$_{\odot}$ and
for illustrative purposes we adopted $\alpha=10^{-2}.$

\begin{figure}
\plotone{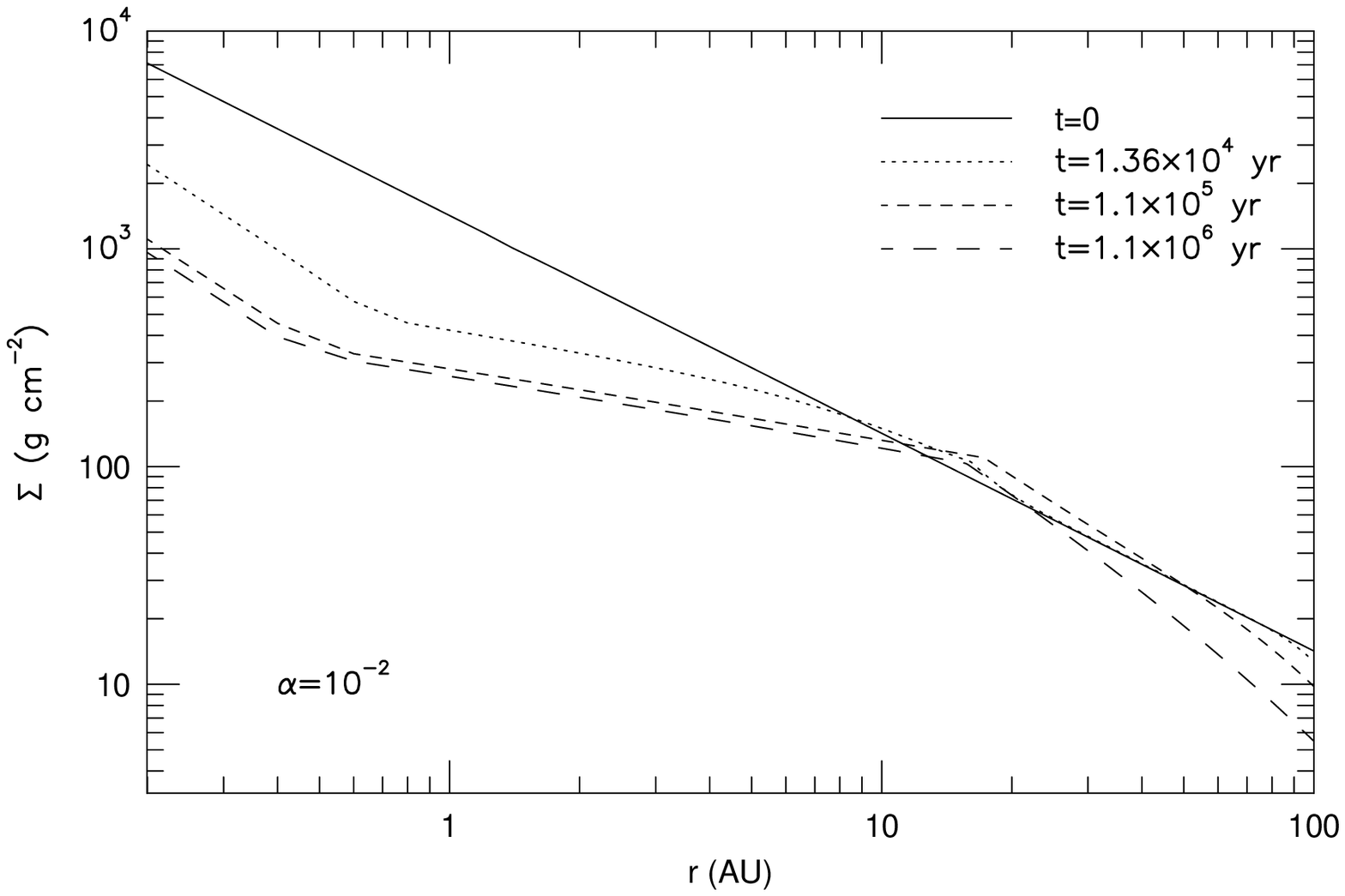}
\caption[]{Solution of the diffusion equation. Shown is $\Sigma$ in
units g~cm$^{-2}$ vs. $r$ in AU using a logarithmic scale plotted at
times $t=0$ ({\it solid line}), $t= 1.36\times 10^4$~yr ({\it dotted
line}), $t= 1.1\times 10^5$~yr ({\it short--dashed line}) and
$t=1.1\times 10^6$~yr ({\it long--dashed line}).  This run has
$\alpha=10^{-2}.$ The total disc mass decreases because of accretion
onto the central object}
\label{fig3}
\end{figure}

In Figure~\ref{fig3} we show the evolution of $\Sigma$ as a function
of time. After a time $\sim 10^6$~yr, the $\Sigma$ profile resembles
that of a steady disc in the inner parts justifying the assumption of
a steady state disc model.  After  this time, the model
is similar to that assumed for the primordial solar nebula with
$\Sigma \sim 200$~g~cm$^{-2}$ at 5~AU.

\section{Planetesimal dynamics}

It is thought that planetesimals can be built up from $\mu {\rm m}$
sized particles through processes of collision, sticking and
accumulation occurring in a gaseous medium with some degree of
turbulence (see the review by Weidenschilling \& Cuzzi 1993).  If
physical parameters are favourable, particles with a size distribution
ranging up to $\sim$~a few km can be produced on timescales on the
order of $10^4$~yr at 1AU. The efficiency of these processes are very
uncertain depending as they do on sticking probabilities and the
degree of particle settling in a turbulent medium etc.  Planetesimal
formation efficiency may be a function of the nebula location, being
more effective beyond a few AU, where ice has condensed.  Here we
assume that planetesimals with mass $m_p \sim 10^{18}$~g may form on a
sufficiently rapid timescale anywhere in the nebula.

We suppose the planetesimals have number distribution $n(m)\propto
m^{-q},$ for some index $q.$ Here the number of planetesimals in the
mass range $(m, m+dm)$ is $n(m)dm.$ Then most of the mass is
distributed in the most/least massive objects according as $ q < 2$ or
$q> 2.$ The surface density of matter in the form of planetesimals is
$\Sigma_p.$ The characteristic number density, $N(m_p),$ for
characteristic mass $m_p$ is then $\Sigma_p/(2m_ph_p).$ Here $h_p$ is
the scale height of the distribution $(\sim \sqrt{2\pi/3} \;
{\overline v}/{\Omega}),$ where ${\overline v}$ is the root mean
square velocity dispersion.

For the typical values, $\Sigma_p=1$~g~cm$^{-2},$ $m_p = 10^{21}$~g,
and $h_p=10^{12}$~cm, the mean distance between planetesimals is $\sim
10^{11}$~cm.  This suggests use of a local box model (e.g. Stewart \&
Wetherill 1988). In this box, the centre of which is in circular orbit
with the local keplerian angular velocity, planetesimals are assumed
to move between encounters with constant velocity.  In this respect,
the effect of the central mass is ignored, and the planetesimals are
treated using the methods of kinetic theory.  The idea (Safronov 1966)
is that the velocity dispersion of the planetesimals is increased by
gravitational scattering, enhancing direct collisions between them
through which they accumulate and grow.

The local box model fails when the largest planetesimals become
isolated.  That happens when accumulation has reached the stage where
there are so few that they are in non--overlapping near circular
orbits such that they cannot perturb similar mass objects in
neighbouring orbits into collision.  In this situation the central
mass can no longer be neglected.  To get an approximate idea of when
isolation occurs at radius $r,$ equate $$\Sigma_p= { m_p\over 8\pi
(m_p/M_{*})^{1/3}r^2}.$$ This is a statement that there is one object
in an annulus of width equal to four times its Roche lobe size. Then
$$m_p=(8\pi \Sigma_p r^2)^{3/2} (M_*)^{-1/2}.$$ This gives $m_p=
10^{27}$~g at 5~AU for $\Sigma_p=1$~g~cm$^{-2}$ and
$M_{*}=1$~M$_{\odot}$ with obvious scalings to other surface densities
and radii.  Thus on the order of 1~M$_{\oplus}$ may be obtained at
5AU.  This argument assumes circular orbits which is probably
reasonable for the largest objects which are circularized through
dynamical friction.  After the isolation stage, planetesimal evolution
is probably best followed by global N--body methods (e.g. Aarseth et
al. 1993). In the gas free case the ultimate result is expected to be
the formation of terrestrial planets.

\subsection{Gravitational scattering and  velocity dispersion}

For a member of a swarm of equal masses interacting gravitationally,
the cross section for elastic scattering is on average larger than
that for a direct collision, so long as the velocity dispersion is
less than the escape velocity. Thus when the velocity dispersion of
the swarm is initially small, it will increase because of the effect
of elastic scatterings.

The interaction radius for two masses $m_p$ to give significant
scattering is $r_{x}= 2Gm_p/{\overline v}^2.$ The time between
encounters is then $t_c= 1/( N(m_p) \pi r_x^2 {\overline v}
\log(\Lambda)).$ Here the $\log(\Lambda) \sim 10$ term accounts for
more distant collisions (Binney \& Tremaine 1987).  Using the above,
one gets $$t_c={{\overline v}^3\over 4\pi N(m_p) G^2 m_p^2 \log
(\Lambda)} = \sqrt{{3\over 2\pi}}{3\over 40\pi^2} \left( {h_p \over r}
\right)^4 {M_*^2\over m_p\Omega\Sigma_p r^2} .$$ For $m_p=10^{20}$~g,
$\Sigma_p=1$~g~cm$^{-2}$ and $M_{*}=1$~M$_{\odot}$, one finds at 1~AU
that $t_c= 2\times 10^{17} (h_p/r)^{4}$~yr.

Note that $h_p$ is small because the planetesimal dispersion
velocities are expected to reach the escape velocity for the
characteristic mass at a maximum.  At this point inelastic physical
impacts become as important as scattering and damp the random motions.
Then for $m_p=10^{20}$~g, ${\overline v}=1.7\times 10^3$~cm~s$^{-1}$,
$h_p/r=8\times 10^{-4}$ and $t_c\sim 9\times 10^4$~yr at 1~AU.

Thus early planetesimal build up occurs on a rapid timescale.

\subsection{Runaway Accretion}

From the above arguments, the collision time of $m_p$ with $m_{p'}$ is
inversely proportional to $N(m_{p'})( m_p+ m_{p'})^2.$ If this does
not decrease with $m_{p'}$, then collisions with larger masses are
dominant and we expect velocity dispersions to build up to the escape
velocity of the largest body (Safronov 1966).  This occurs for $q< 2.$

For $q>2,$ the largest bodies collide with predominantly smaller ones
and are circularized by dynamical friction. They can then accrete
efficiently from the smaller ones which move with a velocity
dispersion small compared to the escape velocity from them.

The accretion rate from planetesimals with mass $m_{p'}$ and velocity
dispersion ${\overline v}_{p'}$ is enhanced by gravitational focusing.
Thus the growth of the largest mass $m_p$ say, with radius $R_p$ is
given by

$${dm_p\over dt}=N(m_{p'})m_{p'}\pi R_p^2{\overline v}_{p'}\left
(1+{2Gm_p\over R_p {\overline v}_{p'}^2}\right).$$ Runaway is caused
both by the increase of $R_p$ with $m_p$ and the gravitational
focusing term in brackets.  The timescale for growth to isolation mass
is comparable to the encounter time, $t_c,$ indicated above. But note
that growth may be slowed down through the effect of encounters
between neighbouring runaways producing an increase in the velocity
dispersion that propagates to all components of the system (Ida \&
Makino 1993)

\section{Giant planet formation} 

After a solid protoplanetary core grows to a critical mass of around a
few $M_{\oplus},$ the surrounding gaseous atmosphere can no longer
grow quasi--statically in mass along with it. A process of collapse
ensues possibly leading to dynamical accretion and mass growth to
values characteristic of giant planets.  We review the theory of this
below.  Such a critical core mass model is supported by the indication
from models of Jupiter that it has a solid core of $\sim
5$--$15$~M$_{\oplus}$ (Podolak et al. 1993).

\subsection{Basic equations governing a protoplanetary envelope}

 Let $\varpi$ be the spherical polar radius in a frame with origin at
the centre of the planet's core. We neglect the rotation of the planet
around both its own spin axis and the disc spin axis. We assume that
the envelope is in hydrostatic equilibrium and spherically symmetric,
so that:

\begin{equation}
\frac{dP}{d \varpi} = - g \rho ,
\label{dpdvarpi}
\end{equation}

\noindent where $g=G M / \varpi^2$ is the acceleration due to gravity,
$M(\varpi)$ being the mass contained in the sphere of radius $\varpi$
(this includes the core mass if $\varpi$ is larger than the core
radius). Mass conservation gives:

\begin{equation}
\frac{dM}{d \varpi} = 4 \pi \varpi^2 \rho.
\label{dmdvarpi}
\end{equation}

The thermodynamic variables in the protoplanet envelope are such that
the equation of state of an ideal gas does not normally apply.  Here
we adopt the state--of--the--art Chabrier et al. (1992) equation of
state for a hydrogen and helium mixture. We fix the abundances of
hydrogen and helium to be 0.7 and 0.28 respectively.

The equation of radiative transport is:

\begin{equation}
\frac{dT}{d \varpi} = \frac{-3 \kappa \rho}{16 \sigma
T^3} \frac{L}{4 \pi \varpi^2} ,
\label{dtdvarpi}
\end{equation}

\noindent where $L$ is the luminosity carried by radiation.  Denoting
 the radiative and adiabatic temperature gradients by $\nabla_{rad}$
 and $\nabla_{ad}$ respectively, we have

\begin{equation}
\nabla_{rad} = \left( \frac{\partial \ln T}{\partial \ln P}
\right)_{rad} = \frac{3 \kappa L_{core} P}{64 \pi
\sigma G M T^4} ,
\label{dTdr_rad}
\end{equation}

\noindent and

\begin{equation}
\nabla_{ad} = \left( \frac{\partial \ln T}{\partial \ln P} \right)_s
,
\end{equation}

\noindent with the subscript $s$ meaning that the derivative has to be
evaluated  at  constant entropy.

We assume that the only energy source comes from the core which
outputs the core luminosity $L_{core}$, given by:

\begin{equation}
L_{core} = \dot{M}_{core} \frac{ G M_{core}}{ r_{core}} ,
\end{equation}

\noindent where $M_{core}$ and $r_{core}$ are respectively the mass
and the radius of the core, and $\dot{M}_{core}$ is the rate of
accretion of planetesimals onto the core. We note that it is customary
to take, instead of $L_{core}$, the luminosity supplied by the
gravitational energy which the planetesimals entering the planet
atmosphere release near the surface of the core (see, e.g., Mizuno
1980; Bodenheimer \& Pollack 1986). However, when the mass of the
atmosphere is small compared to that of the core, these two
luminosities are comparable, providing we take for $\dot{M}_{core}$
the rate of accretion of planetesimals onto the core during the phase
of accretion of the atmosphere and not during the phase of the core
formation.  These two rates may be different if the core has migrated
in the disk before accreting the atmosphere.

If $\nabla_{rad} < \nabla_{ad}$, the medium is convectively stable and
the energy is transported only by radiation. In that case
$L=L_{core}$.

When $\nabla_{rad} > \nabla_{ad}$, some energy is transported by
convection. In that case, $L_{core}= L+L_{conv}$, where $L_{conv}$ is
the convective luminosity. We use the expression for $L_{conv}$ given
by mixing length theory (Cox \& Giuli 1968):

\begin{equation}
L_{conv} = \pi \varpi^2 C_p \Lambda_{ml}^2 \left[ \left( \frac{\partial
T}{\partial \varpi} \right)_s - \left( \frac{\partial T}{\partial
\varpi} \right) \right]^{3/2} \sqrt{ \frac{1}{2} \rho g \left| \left(
\frac{\partial \rho}{\partial T} \right)_P \right| } ,
\end{equation}

\noindent where $\Lambda_{ml}=|\alpha_{ml}P/(dP/d \varpi)|$ is the
mixing length, $\alpha_{ml}$ being a constant of order unity, $\left(
\partial T/\partial \varpi \right)_s = \nabla_{ad} T \left( d \ln P /
d \varpi \right)$, and the subscript $P$ means that the derivative has
to be evaluated for a constant pressure. The quantities $\left(
\partial \rho / \partial T \right)_P$ and $\nabla_{ad}$ are given by
Chabrier et al. (1992).

\subsection{Boundary conditions}

We suppose that the planet core has a uniform mass density
$\rho_{core},$ here taken to be 3.2~g~cm$^{-3}.$ The core radius,
which is the inner boundary of the atmosphere, is then given by:

\begin{equation}
r_{core} = \left( \frac{3 M_{core}}{4 \pi \rho_{core}} \right)^{1/3}.
\end{equation}

The outer boundary of the atmosphere is taken to be
at the Roche lobe radius $r_L$ of
the planet:

\begin{equation}
r_L = \frac{2}{3} \left( \frac{M_{pl}}{3 M_*} \right)^{1/3} r ,
\end{equation}

\noindent where $M_{pl} = M_{core} + M_{atm}$ is the planet mass,
$M_{atm}$ being the mass of the atmosphere, and $r$ is the location of
the planet in the disk (i.e. the separation between the planet and the
central star).

To avoid confusion, we will denote the disk mid--plane temperature,
pressure and mass density at the distance $r$ from the central star by
$T_{mid},$ $P_{mid}$ and $\rho_{mid}$, respectively.

\noindent At $\varpi=r_L,$ the mass is equal to $M_{pl}$, the pressure
is equal to $P_{mid}$ and the temperature is given by:

\begin{equation}
T = \left( T_{mid}^4 + \frac{ \tau_L L_{core}}{4 \pi \sigma r_L^2}
\right)^{1/4},
\end{equation}

\noindent where

\begin{displaymath}
\tau_L = \frac{3}{4} \kappa \left( \rho_{mid}, T_{mid} \right)
\rho_{mid} r_L .
\end{displaymath} 

The condition at $\varpi=r_{core}$ is that the mass is equal to
$M_{core}$ there.

\subsection{Model calculations}

At a given disk radius $r$ and for a given core mass $M_{core},$ we
solve the equations~(\ref{dpdvarpi}), (\ref{dmdvarpi})
and~(\ref{dtdvarpi}) with the boundary conditions described above to
get the structure of the envelope.  The opacity law adopted was the
same as that for the disk models. We note that when the density gets
large, the interior of the envelope becomes convective so that the
value of the opacity does not matter there.

\noindent The equations are integrated using the fifth--order
Runge--Kutta method with adaptive step--size control (Press et
al. 1992).  We guess a starting value of $M_{atm}$ and integrate the
equations from $\varpi=r_L$ down to the core surface
$\varpi=r_{core}$. We then iterate the integration, adjusting
$M_{atm}$ at each step, until the solution gives $M = M_{core}$ at
$\varpi=r_{core}$ with some accuracy.

At each radius $r,$ for a fixed $\dot{M}_{core}$, there is a critical
core mass $M_{crit}$ (which increases as $\dot{M}_{core}$ increases)
above which no solution can be found, i.e. there can be no atmosphere
in hydrostatic and thermal equilibrium confined between the radii
$r_{core}$ and $r_L$ around cores with mass larger than
$M_{crit}$. This is because when the core mass is too large, the
atmosphere has to collapse onto the core in order to supply adequate
luminosity to support itself.  For masses below $M_{crit}$, there are
(at least) two solutions, corresponding to a low--mass and a
high--mass envelope respectively.

\begin{figure}
\plotone{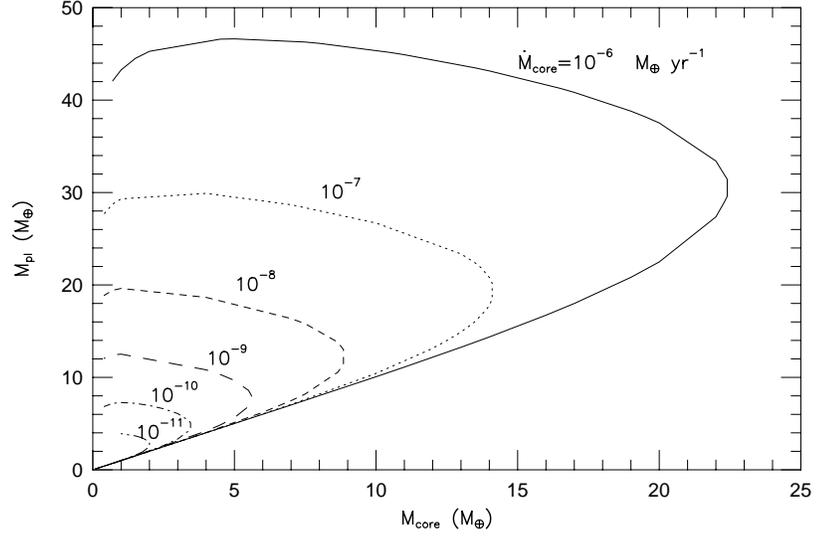}
\caption[]{$M_{pl}$ in M$_{\oplus}$ versus $M_{core}$ in M$_{\oplus}$
for $\dot{M}_{core}$ between $10^{-6}$ and
$10^{-11}$~M$_{\oplus}$~yr$^{-1}$ at a radius $r=5$~AU and for
$T_{mid}=140.05$~K and $P_{mid}=0.13$~dyn~cm$^{-2}$.  From one curve
to another, starting from the right, $\dot{M}_{core}$  decreases
by a factor 10. The critical core mass increases with
$\dot{M}_{core}$, varying between $\sim 2$ and 22.5 M$_{\oplus}$.}
\label{fig4}
\end{figure}

In Figure~\ref{fig4} we plot $M_{pl}$ versus $M_{core}$ for different
$\dot{M}_{core}$ (between $10^{-6}$ and
$10^{-11}$~M$_{\oplus}$~yr$^{-1}$) at a radius of 5~AU and for
$T_{mid}=140.05$~K and $P_{mid}=0.13$~dyn~cm$^{-2}$. These values of
the temperature and pressure are obtained from the vertical structure
integrations described above when the parameters $\alpha=10^{-2}$ and
$\dot{M}_{st} =10^{-7}$~M$_{\odot}$~yr$^{-1}$ are used at $r=5$~AU.

The critical core mass, which decreases when $\dot{M}_{core}$
decreases, is found to be 22.5~M$_{\oplus}$ for $\dot{M}_{core} =
10^{-6}$~M$_{\oplus}$~yr$^{-1}$ and 2.5~M$_{\oplus}$ for
$\dot{M}_{core} = 10^{-11}$~M$_{\oplus}$~yr$^{-1}$. These values are
slightly larger than those found by Bodenheimer \& Pollack
(1986). The difference may be accounted for by the fact that we do not
calculate the luminosity in exactly the same way as they do. Also we
use a slightly different boundary condition for the temperature at the
surface of the planet.

\section{Disc protoplanet interaction}

Once the planetary mass has attained values of around an earth mass or
higher, dynamical interactions with the surrounding disc matter become
important, leading to phenomena such as inward orbital migration and
gap formation (Lin \& Papaloizou 1993; Ward 1997; Lin et al. 1998).

\noindent Korycansky \& Papaloizou (1996) considered the perturbed
disc flow around an embedded protoplanet when the imposed viscosity
$\nu=0.$ They used a shearing sheet approximation in which a patch,
centered on the planet, corotating with its orbit is considered in a
2D approximation.  For unit of length $r_t=r(M_{pl}/M_{*})^{1/3}$ was
adopted, where $r$ is the planet's orbital radius. We remark that $r_t
= (3^{4/3} r_L)/2,$ is a multiple of the Roche lobe radius used above.
When the basic equations are expressed in dimensionless units, the
only parameter defining the problem is ${\cal M}= r_t/(c_s/\Omega),$
being essentially the ratio of Roche lobe radius to disc
semi--thickness.

\begin{figure}   
\rotate[l]{\rotate[l]{
\plotone{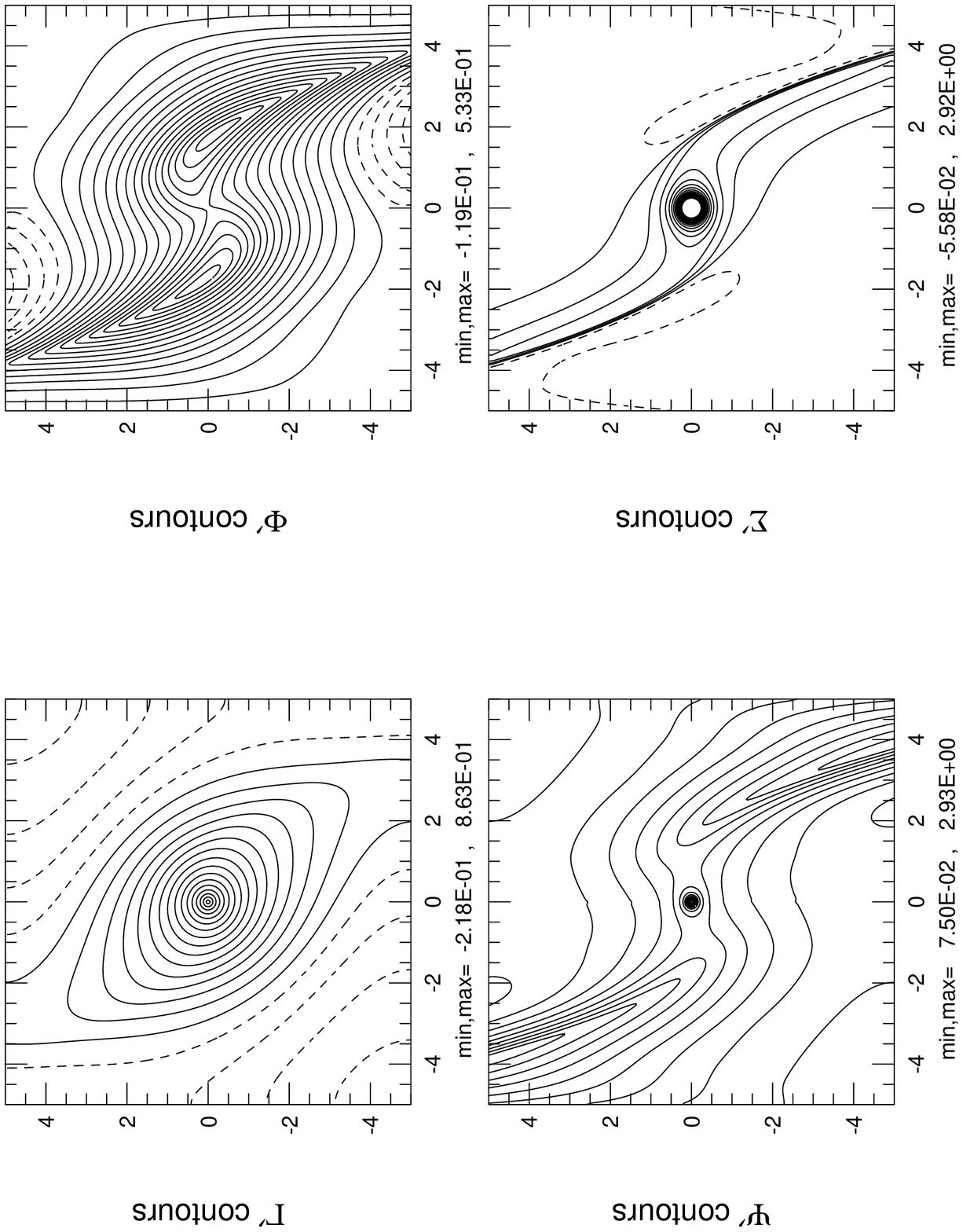}
}}
\caption[]{Flow quantities for the case ${\cal M}=0.7$ taken from
Korycansky \& Papaloizou (1996).  Minimum and maximum contour levels
plotted as indicated in the panels.  Negative contours are indicated
by dashed lines. Top left: Disturbance quantity $\Gamma'$, the
vortical part of the velocity perurbation.  Top right: Disturbance
quantity $\Phi'$, the potential part of the velocity
perturbation. Bottom left: Disturbance quantity $\Psi'$, the momentum
stream function of the perturbed flow. Bottom right: Disturbance
quantity $\Sigma'$, the surface density perturbation}
\label{fig5}
\end{figure}

\begin{figure}
\rotate[u]{
\plotone{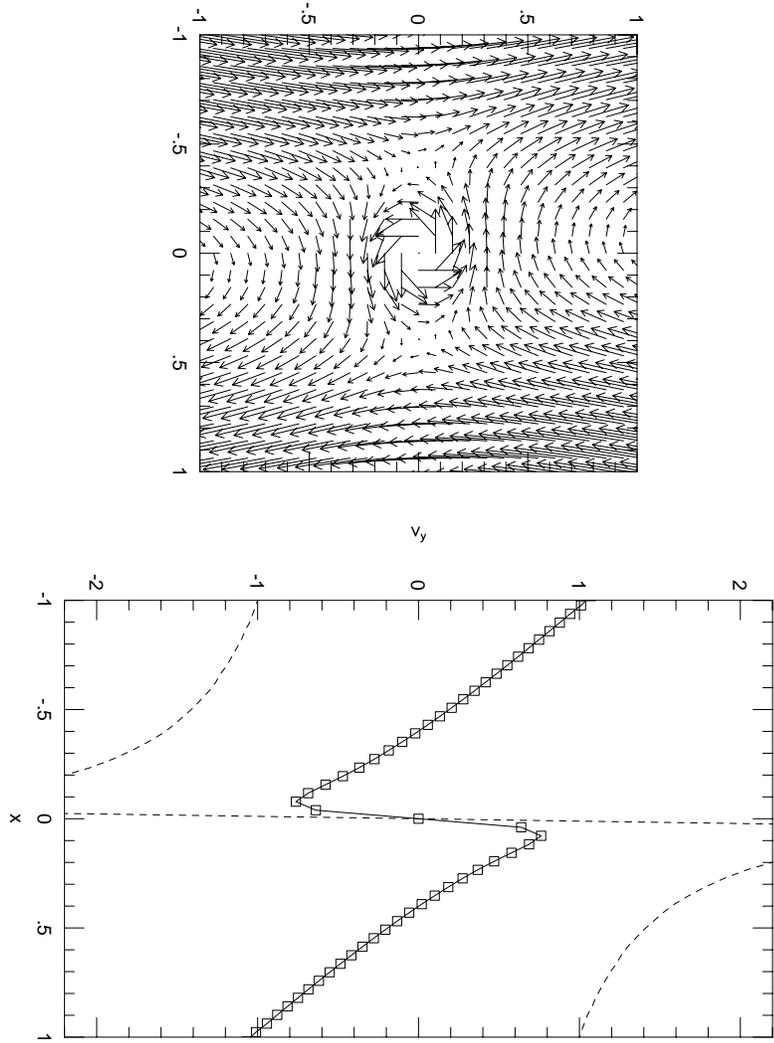}
}
\caption[]
{Velocity field (including background) near the protoplanet taken
from Korycansky \& Papaloizou (1996).
Left: Flow field. 
Right: Velocity $v_y$, along the line
$y=0$. 
\label{fig6}}
\end{figure}

The velocity ${\bf v}$ viewed in the rotating frame
was split  into  components involving a vortical
function $\Gamma$ and a potential $\Phi$:
\begin{equation}
{\bf v} =\nabla\times{\hat z}\Gamma+ \nabla\Phi  \label{eq:dv}
\end{equation}
For a steady flow, there is also a stream function $\Psi,$
such that
\begin{equation}
{\bf v}= \left(\nabla\times{\hat z}\Psi\right)/\Sigma \label{STR}
\end{equation}
Parameters relating to the flow for ${\cal M}=0.7$ are plotted in
figures 5 and 6.  Here Cartesian coordinates are adopted with origin
at the centre of the protoplanet and $x$ axis pointing along the line
joining the central star to the protoplanet.  The perturbed surface
density (figure 5) shows trailing shock waves behind which there is a
strong surface density enhancement giving rise to pronounced wakes.
Also notable is the prograde disc flow around the protoplanet (figure
6).  Dissipation in the shock waves, which become strong once ${\cal
M}> 1,$ eventually results in gap formation (Lin \& Papaloizou 1993).

Protoplanet--disc interaction leads to gap formation and orbital
migration, which, together with tidal interaction with the central
star (Terquem et al. 1998) can lead to massive planets in circular
orbits, with periods of a few days, as observed (Butler et al. 1997).
The reader is referred to the article by Lin et al. (1998) for an
account of the dynamical phenomena which can play a role in
determining the final orbital configuration of planetary systems.

\end{document}